\documentclass[conference,a4paper]{IEEEtran}
\IEEEoverridecommandlockouts
\makeatletter
\newcommand{\linebreakand}{%
  \end{@IEEEauthorhalign}
  \hfill\mbox{}\par
  \mbox{}\hfill\begin{@IEEEauthorhalign}
}
\makeatother
\usepackage{amsmath,amssymb,amsfonts}
\usepackage{algorithmic}
\usepackage{graphicx}
\usepackage{textcomp}
\usepackage{xcolor}
\usepackage{svg}
\usepackage{multirow}
\usepackage{makecell}
\usepackage{amsmath,amssymb,amsfonts}
\usepackage{algorithmic}
\usepackage{graphicx}
\usepackage{textcomp}
\usepackage{float}
\usepackage[caption = false]{subfig}
\usepackage{siunitx}
\usepackage{scalerel}
\usepackage{multirow}
\usepackage{pgfplots}
\usepackage{soul}
\usepackage{acronym}
\pgfplotsset{compat=newest}
\usepackage[center]{caption}
\usepackage{eurosym}
\usepackage{multicol}
\usepackage{booktabs}
\usepackage{svg}
\usepackage{enumitem,kantlipsum}
\usepackage{adjustbox}
\usepackage{acronym}
\usepackage{svg}

\usepackage[T1]{fontenc}
\usepackage{newtxtext,newtxmath} % Font Times-like con supporto matematico

\acrodef{lcoe}[LCOE]{Levelized Cost Of Energy}
\acrodef{dod}[DOD]{Depth Of Discharge}
\acrodef{milp}[MILP]{Mixed Integer Linear Programming}
\acrodef{bess}[BESS]{Battery Energy Storage System}
\acrodef{res}[RES]{Renewable Energy Source}
\acrodef{hess}[HESS]{Hydrogen Energy Storage System}
\acrodef{fc}[FC]{Fuel Cell}
\acrodef{zes}[ZE-Ships]{Zero Emission Ships}
\acrodef{loh}[LOH]{Level Of Hydrogen}
\acrodef{soc}[SoC]{State of Charge}
\acrodef{ci}[CI]{Cold-Ironing}
\acrodef{ems}[EMS]{Energy Management System}
\acrodef{pv}[PV]{Photovoltaic}
\acrodef{wf}[WF]{Wind Farm}
\acrodef{ocv}[OCV]{Open Circuit Voltage}
\acrodef{sos2}[SOS2]{Special Ordered Set of Type 2}
\acrodef{ccus}[CCUS]{Carbon Capture, Utilization and Storage}
\acrodef{ee}[EE]{Energy Efficiency}
\acrodef{imo}[IMO]{International Maritime Organisation}
\acrodef{irena}[IRENA]{International Renewable Energy Agency}
\acrodef{un}[UN]{United Nations}
\acrodef{eu}[EU]{European Union}
\acrodef{sc1}[SC1]{first study case}
\acrodef{sc2}[SC2]{second study case}
\acrodef{smes}[SMES]{Superconducting Magnetic Energy Storage}
\acrodef{sps}[SPS]{Shipboard Power System}
\acrodef{afe}[AFE]{Active Front End}
\acrodef{ppl}[PPL]{pulse Power Load}
\acrodef{msb}[MSB]{Main Switchboard}
\acrodef{emal}[EMAL]{Electromagnetic Aircraft Launch system}
\acrodef{cpl}[CPL]{Constant Power Load}
\acrodef{igbt}[IGBT]{Insulated-Gate Bipolar Transistor}
\acrodef{fmu}[FMU]{Functional Mock-up Unit}
\acrodef{fmi}[FMI]{Functional Mock-up Interface}

\usepackage{acronym}
\acrodef{ppl}[PPL]{Pulse Power Load}
\acrodef{sps}[SPS]{Shipboard Power System}
\acrodef{kpi}[KPI]{Key Performance Indicator}
\acrodef{emal}[EMAL]{ElectroMagnetic Aircraft Luncher}
\acrodef{sg}[SG]{Synchronous Generator}
\acrodef{ess}[ESS]{Energy Storage System}
\acrodef{PHIL}[PHIL]{Power Hardware In the Loop}
\acrodef{SIL}[SIL]{Software In the Loop}
\acrodef{CHIL}[CHIL]{Control Hardware In the Loop}
\acrodef{ems}[EMS]{Energy Management System}
\acrodef{fel}[FEL]{Free-electron Laser}
\acrodef{sc}[SC]{Supercapacitor System}
\acrodef{emr}[EMR]{Electromagnetic Railgun}
\acrodef{epm}[EPM]{Electric Propulsion Motor}
\acrodef{ice}[ICE]{Internal Combustion Engine}
\acrodef{fess}[FESS]{Flywheel Energy Storage System}
\acrodef{bess}[BESS]{Battery Energy Storage System}
\acrodef{pcc}[PCC]{Point of Common Coupling}
\acrodef{csc}[CSC]{Current Source Converter}
\acrodef{vsc}[VSC]{Voltage Source Converter}
\acrodef{pcs}[PCS]{Power Conditioning System}
\acrodef{fmi}[FMI]{Functional Mock-up Interface}
\acrodef{fmu}[FMU]{Functional Mock-up Unit}
\acrodef{srfpll}[SRF-PLL]{Synchronous Reference Frame Phase Locked Loop }
\acrodef{avr}[AVR]{Automatic Voltage Regulator}
\acrodef{itm}[ITM]{Ideal Transformer}
\acrodef{emt}[EMT]{Electromagnetic Transient}

\acrodef{trl}[TRL]{Technology Readiness Level}
\acrodef{ups}[UPS]{Uninterruptible Power Supply}
\acrodef{kpi}[KPI]{Key Performance Indicator}
\acrodef{saidi}[SAIDI]{System Average Interruption Duration Index}
\acrodef{eql}[EqL]{Equivalent Load}
\acrodef{pdf}[pdf]{probability density function}
\acrodef{mttf}[MTTF]{Mean Time To Failure}
\acrodef{mttr}[MTTR]{Mean Time To Repair}
\acrodef{baah}[BAAH]{Breaker-And-A-Half}
\acrodef{dbdb}[DBDB]{Double-Bus-Double-Breaker}
\acrodef{mbt}[MBT]{Manually Bus Transfer}

\usepackage{siunitx} 

\usepackage{color}
\DeclareSIUnit{\kW}{kW}
\DeclareSIUnit{\Wh}{Wh}
\DeclareSIUnit{\MWh}{MWh}
\DeclareSIUnit{\Wp}{Wp}
\DeclareSIUnit{\pu}{p.u.}
\DeclareSIUnit{\EUR}{\mbox{\text{\euro}}}
\DeclareSIUnit{\EURkWh}{\mbox{\text{\euro}}/kWh}
\DeclareSIUnit{\EURkW}{\mbox{\text{\euro}}/kW}
\DeclareSIUnit{\MEUR}{M\mbox{\text{\euro}}}
\DeclareSIUnit{\EURMWh}{\mbox{\text{\euro}}/MWh}

\begin{document}

\title{A Preliminary Assessment of Shipboard Power System Architectures for LVDC Integration}
%\title{Assessing SPS Performance in LVDC Architectures: A Comparative Study }

\author{
\IEEEauthorblockN{D. Roncagliolo}
\IEEEauthorblockA{\textit{University of Genova - DITEN}\\
daniele.roncagliolo@edu.unige.it}
\and
\IEEEauthorblockN{M. Gallo}
\IEEEauthorblockA{\textit{University of Genova - DITEN}\\
marco.gallo@edu.unige.it}
\and
\IEEEauthorblockN{D. Kaza}
\IEEEauthorblockA{\textit{Cetena S.p.A.}\\
daniele.kaza@cetena.it}
\and
\linebreakand
\IEEEauthorblockN{F. D'Agostino}
\IEEEauthorblockA{\textit{University of Genova - DITEN}\\
fabio.dagostino@unige.it}
\and
\IEEEauthorblockN{A. Chiarelli}
\IEEEauthorblockA{\textit{Fincantieri S.p.A.}\\
antonio.chiarelli@fincantieri.it}
\and
\IEEEauthorblockN{F. Silvestro}
\IEEEauthorblockA{\textit{University of Genova - DITEN}\\
federico.silvestro@unige.it}
}

\maketitle

\begin{abstract}
%In the naval sector, a specific challenge lies in managing pulse power loads, whose nature introduces voltage and frequency fluctuations. To mitigate these effects and ensure compliance with standards, energy storage system technologies may be employed. Given the dc nature of most energy storage systems and pulse power loads, the implementation of LVDC sections emerges as a promising solution. References standard such as IEEE Std 45.1-2023 and related literature propose several distribution topologies applicable to LVDC grids. Since the selection of the architecture depends on performance requirements, quantitative analysis is essential. This work compares three notional shipboard power system architectures, incorporating LVDC sections retrofitted from an existing MVAC–LVAC system. The analysis uses five key performance indicators: weight, volume, technology readiness level, system average interruption index, and a pulse power load duration index. SAIDI was originally defined in IEEE Std 1366 for terrestrial distribution networks, with the aim of assessing the average expected hours of downtime for users. This work extends the use of SAIDI to evaluate the impact of distribution topology in a naval context. The results highlight how topology affects overall system performance and the integration of pulse loads.
%%% NEW %%%%
The adoption of low-voltage direct current sections within grid architectures is emerging as a promising design option in the naval sector.
This paper presents a preliminary comparative assessment of three different grid topologies, using an existing MVAC-LVAC shipboard power system as a reference: a conventional MVAC-LVAC radial distribution with an additional LVDC section, a full LVDC radial distribution and a zonal LVDC distribution. Each architecture includes typical elements such as synchronous generators, propulsion motors, energy storage system units, extra propulsive loads, and pulse power loads. The analysis exploits five key performance indicators: weight, volume, technology readiness level, average system interruption duration index, and pulsed power loads interruption index.
\end{abstract}


\section{Introduction}
The naval sector increasingly requires the integration of \acp{ppl}. Due to their intermittent nature, \acp{ppl} introduce repetitive voltage and frequency fluctuations, placing continuous stress on \ac{sps}. To mitigate these disturbances and ensure compliance with standards in the context of power quality requirements (MIL-STD-1399 \cite{MIL_STD_1399_Part_2}, STANAG-1008 \cite{STANAG1008}), \acp{ess} are presented as possible solutions \cite{ESSoverview}. Various \acp{ess} technologies have been proposed, including \acp{bess} in \cite{bessAC} and \cite{bessDC}, \ac{smes} systems in \cite{pplWithSMESS} and \cite{pplWithSMESS2}, \acp{fess} in \cite{fess1} and \cite{fess2}, and rotating mass energy storage for railguns in \cite{Railguns}.

In previous studies, single \acp{ess} units per \ac{ppl} are considered. From a system perspective, when multiple \acp{ppl} have to be integrated, customized solutions can be considered. For example, studies such as \cite{multiplePPLac1}, \cite{multiplePPLac2}, and \cite{multiplePPLdc} propose system-level approaches to accommodate multiple \acp{ppl} by adopting \acp{ess} within specific grid architectures. Considering the dc nature of most \acp{ess} and \acp{ppl}, and recent advances in on-board LVDC systems, the implementation of LVDC sections emerges as a promising design solution \cite{ZonalLVDC}. In this context, IEEE Std 45.1-2023 \cite{ieee45.1} provides design guidelines for \acp{sps} with voltage levels up to 13 kV ac and 1 kV dc. IEEE Std 45.1-2023 introduces electrical power system topologies such as radial and zonal. In the MVDC context, the authors in \cite{DCring} proposed different grid topologies where dc systems are integrated in the \ac{sps}. Additionally, in \cite{DCring-BAAH-DBDB}, a resiliency analysis is performed considering the integration of different grid architectures (e.g., ring bus).

Choosing the most suitable architecture is case-dependent, and a quantitative analysis is more effective than qualitative assessments. A specific performance evaluation of \ac{sps} requires the definition of quantifiable metrics. In \cite{MITcomparisonAnalysis}, configurations are compared using metrics such as weight, volume, fuel consumption, and vulnerability. The study shows that, while the ring topology performs better in terms of vulnerability, it requires more space and additional weight. In \cite{Architectures4ReliabilityImprovement}, six performance indices are used to compare different dc architectures: affordability, efficiency, reliability, survivability, power density, and simplicity of integration. The results highlight benefits and drawbacks between topologies. Zonal systems provide the best resilience, but lower affordability and integration simplicity. In \cite{flexibility}, a quantitative comparison of power distribution topologies based on system flexibility metrics is presented. The study shows that zonal architectures offer the highest distribution flexibility, at the cost of increased system complexity and reduced affordability. On the other hand, radial configurations provide the simplest and most affordable solution, but exhibit the lowest flexibility. Closed ring configuration exhibits balanced performance compared to those studied, providing moderate flexibility and less complexity than zonal systems. The articles compare different architectures by defining and numerically evaluating performance indices. 

This paper presents a comparative analysis of different grid topologies, where an existing MVAC–LVAC \ac{sps} is adopted as the reference case, and the new architectures include LVDC sections. The architectures include the main \ac{sps} elements, such as four \acp{sg}, two \acp{epm}, two \acp{ess}, extra propulsive loads and four LVDC \acp{ppl}. The proposed designs present variations in the distribution system, reflecting alternative approaches to the integration of LVDC technology. The performances are evaluated based on the quantification of five \acp{kpi}, such as weight and volume, \ac{trl}, \ac{saidi}, and \ac{ppl} integrability.

\ac{saidi}, originally defined in IEEE Std 1366 \cite{ieee1366} for terrestrial distribution networks, measures the average system outage duration of the customers per year. This work extends the use of \ac{saidi} to a naval context, evaluating the impact of the distribution grid architecture on the propulsion system, \acp{ppl} and extra propulsive loads.

To complete this analysis, a \ac{ppl} interruption index is used, quantifying the average annual downtime of \acp{ppl}, offering an additional metric to evaluate the performance of different \ac{sps} configurations.

\section{Proposed Architectures}
This section describes the proposed architectures: \textit{A}, \textit{B}, and \textit{C}, designed to integrate LVDC based \acp{ess} and \acp{ppl} into an existing \ac{sps}. Architecture \textit{A} preserves the original radial ac configuration and adds a dedicated LVDC section in a radial layout. Architectures \textit{B} and \textit{C} extend the LVDC approach to the main distribution system, adopting radial and zonal topologies, respectively. The architectures are presented as conceptual layouts and illustrated using single-line diagrams, where the LVDC sections are highlighted in red. Below is a summary of the main \ac{sps} elements:

\begin{itemize}
    \item Four \acp{sg} operating at 6.6 kV ac
    \item Two propulsion systems with a rated voltage of 690 V ac
    \item Two 440 V ac loads, representing the extra propulsive loads on each low-voltage ac  bus.
    \item Two \acp{ess} operating at 1 kV dc
    \item Four \acp{ppl} at 1 kV dc, integrated through a manual transfer switch to provide a backup power supply
\end{itemize}

\subsection{Architecture A, radial ac}
\begin{figure}[h]
    \centering
    \includegraphics[width=0.975\linewidth]{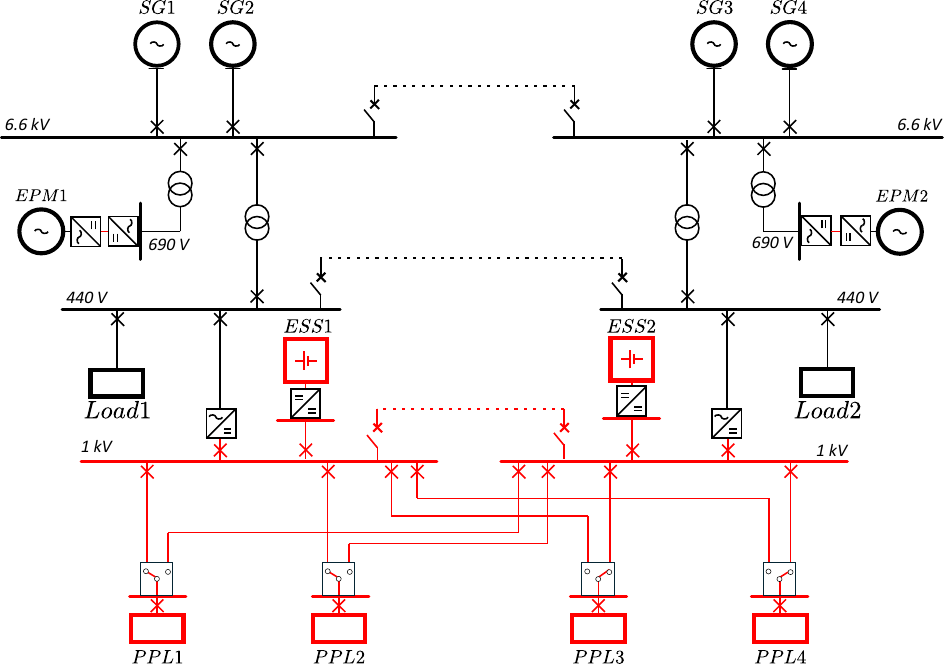}
    \caption{Architecture A, radial ac.}
    \label{fig:Architecture A}
\end{figure}

Architecture \textit{A} is based on a  MVAC-LVAC radial \ac{sps}. An LVDC section is integrated on the low voltage ac side to accommodate \acp{ppl} and \acp{ess}. Figure~\ref{fig:Architecture A} shows the proposed configuration.

The \ac{sps} is divided into three sections. 
The MVAC section features a radial topology with two \acp{msb} at 6.6 kV ac. It includes the power station, the propulsion system, and the distribution transformers to connect the low voltage side. The \acp{epm} are connected to the \acp{msb} through a driver, a \ac{vsc} and a transformer 6.6/0.69 kV/kV ac.
The low voltage side comprises both the LVAC and LVDC sections.
The LVAC section follows a radial topology with two \acp{msb} at 440 V ac, supplying the LVDC section and the extra propulsive loads. The extra propulsive loads are modeled as two equivalent loads: $Load1$ and $Load2$. 
The LVDC section has a radial configuration with two \acp{msb} at 1 kV dc. Each \ac{msb} is connected to the low-voltage ac section via a \ac{vsc}. This section includes an \ac{ess} and two \acp{ppl} per \ac{msb}. Each \ac{ess} consists of two units connected in parallel, assumed to be installed in the same compartment. The \acp{msb} are interconnected by tie-breakers normally open to ensure operation with separated power plants.
This architecture introduces a basic integration of dc elements from a system perspective, while preserving the traditional ac-based structure and maintaining high \ac{trl}.

\subsection{Architecture B, radial dc}
\begin{figure}[h]
    \centering
    \includegraphics[width=0.975\linewidth]{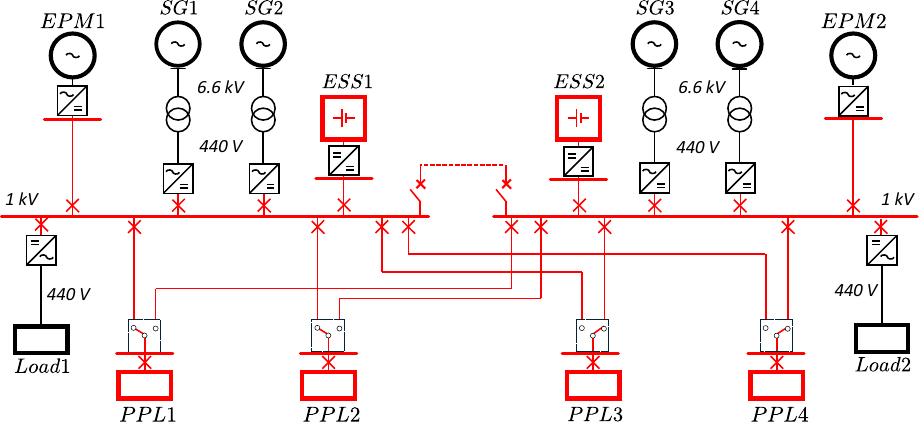}
    \caption{Architecture B, radial dc.}
    \label{fig:Architecture B}
\end{figure}

Architecture B adopts a dc radial configuration. It consists of a predominantly LVDC distribution with two \acp{msb} in dc. Figure~\ref{fig:Architecture B} illustrates the layout.

The \ac{sps} employ a centralized configuration, where the \acp{msb} serve as interconnection points for the main elements of the system. The \acp{msb} are interconnected by normal open tie-breakers to ensure operation with separated power plants.
The adoption of a full dc distribution system requires the integration of transformers and \acp{vsc} to enable compatibility with medium voltage \acp{sg}. The dc stages of the \acp{epm} drivers are directly connected to the dc \acp{msb}. Extra propulsive loads, being ac, require dedicated \acp{vsc} for integration.
Each dc \ac{msb} is connected to an \ac{ess} and two \acp{ppl} per \ac{msb}. Each \ac{ess} consists of two units connected in parallel, assumed to be installed in the same compartment.
%Although the architecture reduces the number of \acp{msb}, saving weight and space, it results in a reduced overall system reliability.

\subsection{Architecture C, zonal dc}
\begin{figure}[h]
    \centering
    \includegraphics[width=0.975\linewidth]{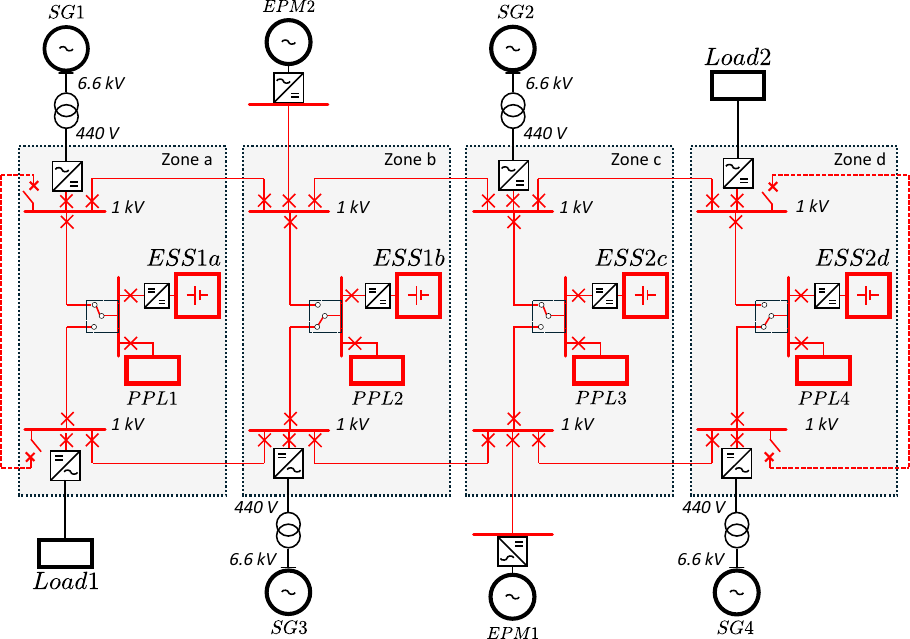}
    \caption{Architecture C, zonal dc.}
    \label{fig:Architecture C}
\end{figure}

Architecture C is based on the dc zonal configuration. This topology uses the concept of electrical zones that are designed to operate independently. Figure~\ref{fig:Architecture C} presents the configuration.

The \ac{sps}, based on a 1 kV dc ring topology, is divided into four electrical zones. Each zone receives input from one \ac{sg} via a transformer and a \ac{vsc}, ensuring electrical independence between zones. The ring is electrically accessible through two \acp{msb} per zone.

\ac{epm}1 and \ac{epm}2 are connected to zone b and zone c, respectively. The dc stages of their drivers are directly connected to the dc \acp{msb}. The two \acp{ess} of the previous architectures are split into four units. Each unit is assigned to a dedicated bus that supplies the \acp{ppl} in each zone. extra propulsive loads are connected to zone a and zone d with \acp{vsc}.

The \acp{msb} on the extreme sides of the loop are coupled with normal open tie-breakers.

%This architecture improves system redundancy. Although it increases overall volume and weight, its spatial distribution can reduce the cable length.

\section{Availability estimation}
This section presents the methodology used to estimate the availability of the main \ac{sps} elements for each configuration. Fault events are attributed to the electrical aging of transformers, lines, switchboards, converters and breakers. The analysis disregards the intrinsic reliability of the system components themselves and considers only the elements of the network infrastructure that enable the connection to the grid. The assumed failure rate ($\lambda_i$) and Mean Time To Repair ($MTTR_i$) values are shown in Table~\ref{tab:reliability_index} and are taken from \cite{Architectures4ReliabilityImprovement,TransfRelIndex,MarkovStudycase1}.

\renewcommand{\arraystretch}{1.5} % Aumenta altezza delle righe
\begin{table}[h]
    \centering
    \caption{Component reliability index}
    \label{tab:reliability_index}
    \renewcommand{\arraystretch}{2} % aumenta l'altezza delle righe del 50%
    \begin{tabular}{|c|c|c|}
    \hline
    \makecell{{\textbf{Infrastructure}}\\{\textbf{Element}}} & \makecell{{$\boldsymbol{\lambda_i}$}\\{[failures/year]}} & \makecell{{$\mathbf{MTTR_i}$}\\{[hours]}}\\
    \hline
    Transformer & 0.012 & 168 \\
    \hline
    Line & 0.001 $L$ & 8 \\
    \hline
    Bus & 0.01 & 8 \\
    \hline
    Converter & 0.006 & 1 \\
    \hline
    Breaker & 0.01 & 4 \\
    \hline
    \end{tabular}
\end{table}

\renewcommand{\arraystretch}{1}
Here, $L$ represents the length of the line in meters. Line lengths are based on those of a real \ac{sps}, with adjustments made for the dc configurations and the specific subsystems considered. The failure rate and repair rate associated with each $i$-th item are defined as $\lambda_i = \frac{1}{\text{MTTF}_i}$ and $\mu_i = \frac{1}{\text{MTTR}_i}$, respectively.

%In addition, the lengths for each subsystem have been taken into account.

%Here, $L$ represents the length of the line in meters. The lines length are based on the length of a real \ac{sps} with the correction made for the dc configurations. In addition, for each subsystem, the lenght have been considered.

%Since differences in cable length can affect the reliability of the system \cite{CableImpact}, it is assumed that the line failure rate is linearly proportional to its length. In this way, the spatial distribution of the electrical infrastructure can influence the general availability of the system. 

\subsection{Method and Assumptions}
A Markovian analysis is used to evaluate the system availability, with the following assumptions adopted:

\begin{itemize}
    %\item Non-simultaneity of faults and repairs
    \item The faults and repairs are Mutually Exclusive and Collectively Exhaustive;
    \item Each fault event is statistically independent;
    \item Faults are mutually exclusive in time;
    \item Transfer switches operate in 30 minutes;
    \item Automated tie-breakers operate in 1 minute.
\end{itemize}
 
These assumptions enable the use of a series system availability model to evaluate the availability of the main \ac{sps} elements.
% Figure~\ref{fig:markov_serie} shows the graphical representation of the Markov chain for a series system with \textit{n} items.
% \begin{figure}[H]
%     \centering
%     \includegraphics[scale=0.4]{Figures/MarkovSerie.pdf}
%     \caption{Markov chain of a series system}
%     \label{fig:markov_serie}
% \end{figure}
% Where $ \lambda_i = \frac{1}{\text{MTTF}_i} $ and $ \mu_i = \frac{1}{\text{MTTR}_i} $ are the failure rate and the repair rate associated with the element $ \text{item}_i $.

The availability of a series system with $n$ items is estimated using the Markov chain theory \cite{trivedi2001}:
\begin{equation}
    A^{sys} = \frac{\mu^{eq}}{\lambda^{eq} + \mu^{eq}} \label{Aeq}
\end{equation}
where $\lambda^{eq}$ and $\mu^{eq}$ result in the equivalent failure and repair rate of the series system. 
$\lambda^{eq}$ is calculated as the sum of the failure rates of the elements.
\begin{equation}
    \lambda^{eq} = \sum_{i=1}^{n} \lambda_i
\label{LAMBDAeq}
\end{equation}
where $\mu^{eq}$ is calculated as the reciprocal of the equivalent main time to repair of the system ${MTTR}^{eq}$. The ${MTTR}^{eq}$ is calculated as the average of the individual items mean time to repair ${MTTR}_{i}$, weighted by the respective failure rates $\lambda_{i}$.
\begin{equation}
MTTR^{eq} = \frac{\sum_{i=1}^{n} \lambda_i \, MTTR_i}{\sum_{i=1}^{n} \lambda_i}
\label{MTTReq}
\end{equation}
\subsection{Availability of the main \ac{sps} elements}

For each configuration, the connection of the main \ac{sps} elements depends on a specific set of components of the grid infrastructure that form a series reliability model. Equations \eqref{Aeq}, \eqref{LAMBDAeq} and \eqref{MTTReq} are applied to each reliability chain associated with the main \ac{sps} elements, and the annual downtime of each item is derived from the resulting availability. 

The main elements are grouped into subsystems: ac generation, propulsion, extra propulsive loads, pulse power loads, and dc generation. In each subsystem, the mean of the results of its elements is used to determine the equivalent failure rate of the subsystem, repair time, and annual downtime.  For each subsystem and configuration, Table~\ref{tab:reliability_series} reports the results.

\renewcommand{\arraystretch}{1.4} % aumenta l'altezza delle righe del 50%
\begin{table}[H]
\centering
\caption{Reliability indices of SPS subsystems}
\label{tab:reliability_series}
\renewcommand{\arraystretch}{2} % aumenta l'altezza delle righe del 50%
\begin{tabular}{|l|c|c|c|}
\hline
\multicolumn{4}{|c|}{\textbf{Architecture A}} \\
\hline
\makecell{{\textbf{Subsystem}}} & \makecell{{$\boldsymbol{\lambda^{eq}}$}\\{[failures/year]}} & \makecell{{$\mathbf{MTTR^{eq}}$}\\{[hours]}} & \makecell{{\textbf{Annual}}\\{\textbf{Downtime}}\\{[hours per year]}}\\
\hline
AC generation      & 0.049 & 16.026 & 0.781 \\ 
Propulsion         & 0.094 & 35.085 & 3.297 \\ 
Extra propulsive loads     & 0.020 & 6.000 & 0.120 \\ 
Pulse power loads  & 0.060 & 2.333 & 0.140 \\ 
DC generation      & 0.041 & 11.488 & 0.471 \\ 
\hline
\hline
\multicolumn{4}{|c|}{\textbf{Architecture B}} \\
\hline
\makecell{{\textbf{Subsystem}}} & \makecell{{$\boldsymbol{\lambda^{eq}}$}\\{[failures/year]}} & \makecell{{$\mathbf{MTTR^{eq}}$}\\{[hours]}} & \makecell{{\textbf{Annual}}\\{\textbf{Downtime}}\\{[hours per year]}}\\
\hline
AC generation      & 0.067 & 41.996 & 2.802 \\ 
Propulsion         & 0.076 & 16.789 & 1.276 \\ 
Extra propulsive loads     & 0.026 & 4.846 & 0.126 \\ 
Pulse power loads  & 0.060 & 2.333 & 0.140 \\ 
DC generation      & 0.041 & 11.488 & 0.471 \\ 
\hline
\hline
\multicolumn{4}{|c|}{\textbf{Architecture C}} \\
\hline
\makecell{{\textbf{Subsystem}}} & \makecell{{$\boldsymbol{\lambda^{eq}}$}\\{[failures/year]}} & \makecell{{$\mathbf{MTTR^{eq}}$}\\{[hours]}} & \makecell{{\textbf{Annual}}\\{\textbf{Downtime}}\\{[hours per year]}}\\
\hline
AC generation      & 0.048 & 49.417 & 2.371 \\ 
Propulsion         & 0.051 & 13.745 & 0.701 \\ 
Extra propulsive loads     & 0.026 & 4.846 & 0.126 \\ 
Pulse power loads  & 0.020 & 6.000 & 0.120 \\ 
DC generation      & 0.026 & 4.846 & 0.126 \\ 
\hline
\end{tabular}
\end{table}

Compared with architecture A, architectures B and C introduce the transformer and converter as additional series elements for synchronous generators, resulting in a higher mean downtime. In architectures B and C, EPM drives are directly connected to dc \acp{msb}, eliminating one conversion stage and thus improving the availability of the propulsion system. The main reliability advantage of architecture C lies in placing an \ac{ess} close to each \ac{ppl}, thus removing their dependency on the line and the \acp{msb} that connect them to the main network. This distributed architecture also reduces the average cable length, improving reliability, as well as reducing weight and volume. In Architectures A and B, the connection of \acp{ppl} additionally depends on the upstream line and the associated \ac{msb}. Since these elements can be bypassed by the operation of a transfer switch, their MTTR is assumed to be equal to the intervention time of the transfer switch.

\section{Comparison of Architecture Performance}
The proposed \acp{sps} present different features, which can be advantageous or disadvantageous. These aspects have been discussed in the sections dedicated to the individual architectures, mainly in a qualitative way. Based on these considerations, the choice of one architecture over another could be made qualitatively, relying on common sense arguments. However, in order to support this process with a more rigorous assessment, a quantitative evaluation is required. For this purpose, the main characteristics are quantified by defining five \acp{kpi}, which are then used to compare the three configurations.

\subsection{\acp{kpi} Definition}
The analysis relies on five \acp{kpi} defined as follows:
% \begin{itemize}
%     \item Compactness through weight and volume calculations
%     \item Availability through the evaluation of average system downtime
%     \item Feasibility through the assessment of the overall \ac{trl}
% \end{itemize}

%The individual \acp{kpi} are defined as follows:

\begin{enumerate}
    \item Weight and volume. 
    The calculation of weight and volume considers the main elements of the grid, such as the generators, \ac{epm}, \ac{ess} and \ac{ppl}. For each component requiring a converter, a switchboard containing all associated elements, including filters, dc breakers and cooling units are considered. Transformers and lines are also included.
    All values are based on commercial components, ensuring that the assessment reflects realistic system implementations while preserving comparability across architectures.

    \item \ac{saidi}.  
    This index, defined in IEEE Std 1366-2012, represents the total duration of interruption for the average customer during a given period. Although traditionally applied to terrestrial power distribution systems, it is here adapted to naval context, where subsystems such as propulsion, \acp{ppl} and extra propulsive loads are considered equivalent to customers. \ac{saidi} is therefore repurposed to measure the average downtime of these subsystems, providing a system-level availability indicator. Taking a year as the reference period, and using the annual downtime of the main \ac{sps} elements calculated in this study, the \ac{saidi} for each configuration is evaluated using equation \eqref{SAIDI}.

    \begin{equation}
        SAIDI [h] = \frac{\sum \text{Annual Downtime}}{N}
        \label{SAIDI}
    \end{equation}
    Where $N$ represents the number of customers. In this work, the number of subsystems considered is $N=3$, which corresponds to propulsion, extra-propulsive loads, and pulse power load.
    \item \ac{ppl} Interruption index.  
    This metric focuses specifically on pulse power loads, representing their average annual downtime. It serves as an indicator of the integrability of \acp{ppl} into the different architectures and quantifies the impact of system topology on their operational availability.

    \item \ac{trl}. 
    The overall TRL for each architecture takes into account the maturity and limitations of commercially available components for the integration, protection, and realization of switchgear and controlgear assemblies. Some components, such as converters, have the highest TRL, while several other components are still in the validation stages. Depending on the type of configuration, the TRL estimate ranges from TRL 5 to TRL 7.
        
\end{enumerate}

\subsection{Results discussion}
Table~\ref{tab:KPItab} summarizes the results for all defined \acp{kpi}. SAIDI and PPL Interruption index is calculated using the methodology proposed in this work, while weight, volume, and TRL are derived from commercially available technologies.

\begin{table}[H]
\centering
\caption{KPIs results comparison}
\label{tab:KPItab}

% Riduce leggermente lo spazio tra le colonne
\setlength{\tabcolsep}{3.5pt} % default è 6pt
\renewcommand{\arraystretch}{2.5} % aumenta l'altezza delle righe del 50%
\begin{tabular}{|c|c|c|c|c|c|}
\hline
\textbf{Conf.} & 
\makecell{\textbf{Weight}\\{[t]}} & 
\makecell{\textbf{Volume}\\{[m$^3$]}} & 
\makecell{\textbf{SAIDI}\\{[hours/year]}} & 
\makecell{\textbf{PPL interruption index}\\{[hours/year]}} & 
\makecell{\textbf{TRL}\\{[-]}} \\
\hline
A & \underline{748} & \underline{589} & 1.186 & 0.140 & \underline{7} \\
B & 771 & 733 & 0.514 & 0.140 & 6 \\
C & 762 & 612 & \underline{0.316} & \underline{0.120} & 5 \\
\hline
\end{tabular}
\end{table}

\begin{figure}[H]
    \centering
     \includegraphics[width=0.975\linewidth]{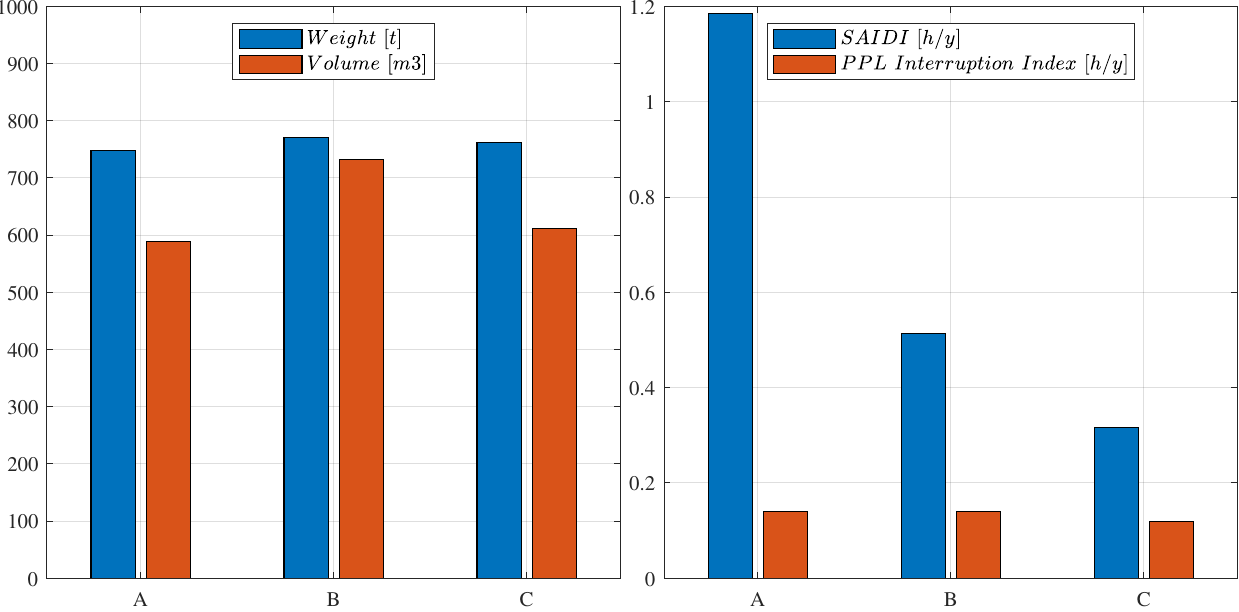}
    \caption{KPI Bar-plot Diagram}
    \label{fig:KPIfig}
\end{figure}

Architecture A is the lightest and most compact solution, achieving the highest TRL. However, due to the absence of standardized selectivity analysis and regulations, its TRL is 7.
Architecture B adopts a centralized dc radial configuration, offering moderate reliability improvements for the EPM subsystem, but incurring additional conversion stages for extra propulsive loads.
Architecture C, based on a DC zonal configuration, enhances system availability and PPL integration by leveraging redundancy, distributed ESS placement, and shorter overall cable runs. This comes at the expense of higher weight, volume, and design complexity. Nevertheless, it achieves the lowest SAIDI and PPL interruption index values, indicating higher reliability performance.
Figure~\ref{fig:KPIfig} provides a visual comparison of the KPIs.
% Red values highlight the best-performing KPI among the configurations.  
% Configuration A is the lightest and most compact (columns 1–2), while C achieves the lowest SAIDI and PPL Interruption (columns 3–4). \textcolor{red}{Configurations A and B have similar reliability indices}. Differences related to propulsion and ac generation system availability are reported in Table~\ref{tab:reliability_series}.
% TRL decreases from Configuration A to C (column 5). Configuration A has the highest TRL due to fully available components, though its TRL is limited to 7 by the lack of standardized selectivity analysis and regulations.  
% To provide a visual comparison of the key performance indicators, a bar-plot diagram is presented in 
%Figure~\ref{fig:KPIfig}, showing weight and volume alongside SAIDI and PPL Interruption for all configurations. 
%This allows a rapid comparison between compactness and reliability performance.

Overall, no single architecture outperforms the others across all KPIs. Architecture A offers the most mature and compact solution, architecture C delivers the highest reliability, while architecture B represents an intermediate option with partial benefits but the least compactness.

% \subsection{Results discussion}
% This paper has presented three notional \ac{sps} architectures with LVDC sections, derived as retrofits from an existing MVAC–LVAC platform.
% Architecture A present a traditional ac radial scheme with an added LVDC section. It provides the highest compactness and TRL, since all components are commercially available.
% Architecture B is based on a centralized dc radial scheme. It reduces weight and space for EPM and improve their availability, but decreases reliability for ac loads and ac generation system due to the requires of additional conversion stages.
% Architecture C adopt the concept of zonal configuration applied on an LVDC system. It improves availability and PPL integration through redundancy and distributed ESS, at the expense of higher weight, volume, and complexity.
% The KPI-based analysis shows that Architecture C achieves the lowest downtime in terms of SAIDI and PPL interruption indices, while Architecture A remains the most mature and compact solution. Architecture B represents an intermediate solution, with partial benefits but it turns out to be the least compact.
% Overall, the results confirm that no single architecture is the best considering all performance indices. 

%The choice of topology should therefore be driven by mission requirements and design priorities, balancing availability, compactness, and technological maturity.
%Future work will focus on the implementation of a dynamic model to further validate and extend the comparative evaluation.

\section{Conclusions}
This study presents a preliminary comparative assessment of three SPS architectures designed to integrate the LVDC distribution for the accommodation of ESS and PPL. The configurations considered include MVAC–LVAC radial distribution with an LVDC section, LVDC radial distribution and LVDC zonal distribution. To evaluate these architectures, five KPIs are analyzed: weight, volume, TRL, SAIDI, and PPL interruption index.

All indicators are evaluated considering the expected physical distribution of the components on board, including converters, filters, and cable lengths. The results reveal that no single architecture excels in all performance metrics. However, some general conclusions can be drawn. The MVAC–LVAC radial distribution offers the highest performance in terms of weight, volume, and TRL, mainly due to the maturity of commercially available components. The LVDC zonal distribution achieves higher availability due to its inherent redundancy and distributed architecture, although it introduces greater complexity, weight, and volume compared to architecture A. The LVDC radial configuration provides an intermediate level of reliability, but represents the least favorable option in terms of weight and volume.

Despite these differences, all configurations lack advanced selectivity analysis and regulatory standardization. Future work will focus on developing both static and dynamic models to validate and extend the analysis, incorporating system-level insights for a more comprehensive evaluation.

\bibliographystyle{IEEEtran} 
\bibliography{biblio.bib}
\end{document}